\documentclass[letterpaper, 10 pt, conference]{ieeeconf}  

\IEEEoverridecommandlockouts                              

\usepackage{graphics} 
\usepackage{epsfig} 
\usepackage{amsmath} 
\usepackage{amssymb}  
\usepackage[pass]{geometry} 

\usepackage[utf8]{inputenc}
\usepackage{t1enc}
\usepackage{bm}
\usepackage{color}
\usepackage{cases}

\newcommand{\figref}[1]{Fig.~\ref{#1}}
\title{\LARGE \bf
Self-organized UAV Traffic in Realistic Environments*
}

\author{Csaba Virágh$^{1}$, Máté Nagy$^{2,3}$, Carlos Gershenson$^{4}$, Gábor Vásárhelyi$^{2}$
\thanks{*This work was supported by the János Bolyai Research Scholarship of the Hungarian Academy of Sciences BO/00219/15/6}
\thanks{$^{1}$Cs.V. is with the Biological Physics Department,
        Eötvös University, Budapest, Hungary
        {\tt\small viraghcs@hal.elte.hu}}%
\thanks{$^{2}$M.N. and G.V. Authors are with MTA-ELTE Statistical and Biological Physics Research Group,
        Budapest, Hungary
        {\tt\small nagymate@hal.elte.hu, vasarhelyi@hal.elte.hu}}%
\thanks{$^{3}$M.N. is with Max Planck Institute for Ornithology, Department of Collective Behaviour \&
	Chair of Biodiversity and Collective Behaviour, University of Konstanz,
	Konstanz, Germany
        {\tt\small mnagy@orn.mpg.de }}%
\thanks{$^{4}$C.G is with IIMAS \& C3, UNAM, Mexico City, Mexico; SENSEable City Lab, MIT, Cambridge, USA; MOBSLab, Northeastern University, Boston, USA \& ITMO University, St. Petersburg, Russian Federation
        {\tt\small cgg@unam.mx }}%
}

\begin{document}

\maketitle
\thispagestyle{empty}
\pagestyle{empty}


\begin{abstract}

We investigated different dense multirotor UAV traffic simulation scenarios in open 2D and 3D space, under realistic environments with the presence of sensor noise, communication delay, limited communication range, limited sensor update rate and finite inertia. We implemented two fundamental self-organized algorithms: one with constant direction and one with constant velocity preference to reach a desired target. We performed evolutionary optimization on both algorithms in five basic traffic scenarios and tested the optimized algorithms under different vehicle densities. We provide optimal algorithm and parameter selection criteria and compare the maximal flux and collision risk of each solution and situation. We found that i) different scenarios and densities require different algorithmic approaches, i.e., UAVs have to behave differently in sparse and dense environments or when they have common or different targets; ii) a slower-is-faster effect is implicitly present in our models, i.e., the maximal flux is achieved at densities where the average speed is far from maximal; iii) communication delay is the most severe destabilizing environmental condition that has a fundamental effect on performance and needs to be taken into account when designing algorithms to be used in real life.

\end{abstract}

\section{INTRODUCTION}

As more and more individual UAVs are present in the common airspace, there is an urgent need for both centralized and decentralized solutions that assure the safety of flying objects when they meet each other in the air. In the long term, three-dimensional air traffic might be as populated and dense as current road traffic. However, current working solutions for road and air traffic control are not designed and neither are suitable for handling a large amount of vehicles in 3D open space. Road traffic control handles millions of vehicles but is highly constrained by one-dimensional lanes that are evident and visible for all cars using them. Centralized traffic control elements (traffic lights, traffic signs etc.) are also of great help to assure safe and maximal autonomous flow of cars with or without drivers in every possible (fixed) junction. However, the air is free from any traffic signs or visible road markers and is three dimensional, which results in fundamentally different and more complex situations when interference of trajectories occurs. Moreover, dense UAV traffic opens new possibilities for vehicle coordination that have not been explored, as most traffic research is constrained to roads~\cite{may1990traffic,helbing2001traffic}, while air traffic control is focused mainly on aircraft with high velocity and inertia at low spatial density~\cite{kovsecka1997generation,tomlin1998conflict,menon1999optimal,5203698}.

Current air traffic control is overwhelmingly centralized and thus limited in terms of scalability~\cite{turpin2013goal,7128399}. With the current structure of air traffic control we will most probably not be able to handle hundred or thousand times more UAVs. Therefore, decentralized algorithms will be necessary to handle local encounters and collision avoidance~\cite{Physicomimetics2012,panagou2014decentralized}. There are more and more UAVs equipped with individual collision avoidance mechanisms but if these are not harmonized and not tested in dense traffic situations, they will surely fail, such as even people in panic situations~\cite{helbing2000simulating} or cars in delay-induced ghost traffic jams and road accidents~\cite{orosz2010traffic}.

Self-organizing strategies have proven to be useful in simulations to coordinate road traffic~\cite{Gershenson2005,Lammer:2008,gershenson2012self}. In general, traffic flow changes constantly thus it is more efficient to regulate it with distributed adaptation of current flow instead of the centralized optimization of an averaged flow.

In this paper we present decentralized, self-organized solutions for dense UAV traffic under realistic conditions. By decentralization we mean that every agent (UAV in the simulation) calculates its own desired outputs locally, based on the local information available to it at a given moment, without central processing of any dynamic global information. Note that our interactions between agents are always local - they are limited both by finite communication range and localised interaction terms, which enables the scalability of the system. By self-organization we mean that any global ``what'' (fleet-level task) is solved by local ``how''. Note that for this we use the same global dynamic equations by all agents; however, locality will be provided by the non-uniform spatial distribution of information available to agents.

We present two distinct models, basically for multirotor type agents that can go in any direction freely and are also capable of hovering in the air. However, one model is based on a constant velocity assumption which can be extended to be used also with fixed-wing aircraft later on. These algorithms have been developed and tested in a simulation framework~\cite{viragh2014flocking} which is capable of taking into account several specific features of autonomous UAVs, such as inaccuracy of the sensors, time delay of the communication between the robots, outer noises or inertial effects. It is often a very hard task to develop and fine-tune an algorithm under these “realistic” conditions, therefore, in contrast to simple physical models, such as the Vicsek-model~\cite{vicsek1995novel} that we incorporate into our solution to some extent, our traffic algorithms contain a larger number of parameters.

In the next section, we introduce the reader to the scenarios which have been studied, the interaction terms which are common in the two traffic algorithms, then, we summarize the unique terms of each algorithm. Note that in the context of this paper, “algorithm” only means that we define specific rules of motion (more precisely: specific desired output velocities) for specific situations and environmental conditions. For more details of the simulation framework and the underlying differential equations, see~\cite{viragh2014flocking}.

\section{THE MODEL}
\subsection{Basic Traffic Scenarios}

\begin{figure}[thpb]
	\centering
	\includegraphics[scale=1.0]{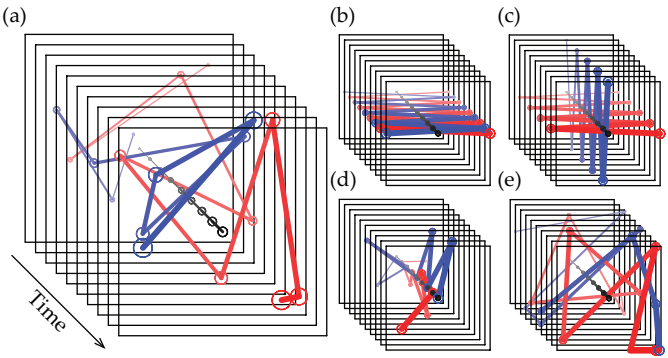}
        \caption{Illustration of the time evolution of the five basic traffic scenarios in 2D: a) random; b) line; c) cross; d) star; e) edge. The red and blue lines correspond to the idealized trajectories of two independent agents between their target points which are shown by circles. The brightness, the size of the target circles and line width of the trajectories indicate time. Each black frame represents a time instant when a target is reached. Note that the random, star and edge scenarios are different in 2D and 3D, while the line and cross remain the same, with an extra degree of freedom for motion in 3D.}
	\label{fig:scenarios}
\end{figure}

In our test cases we simulate traffic by assigning a target point to each agent in a common open space, without any obstacles (if needed, simple obstacles or walls can be modelled as virtual/shill agents, as in~\cite{han2006soft} or~\cite{viragh2014flocking}; however, handling obstacles is not our goal in this context). Every time a target point is reached by an agent, a new target point is assigned. The term “scenario” refers to the actual spatial distribution of the target points. We investigated five basic traffic scenarios with the following target point locations:

\begin{itemize}
	\item RANDOM – random distribution inside a square (2D) / cube (3D) (\figref{fig:scenarios}a);
	\item	LINE – endpoints of a single line, with each agent alternating on the two endpoints (\figref{fig:scenarios}b);
	\item	CROSS – endpoints of two identical, perpendicular lines crossing each other at the middle, with each agent alternating on endpoints of only one line (\figref{fig:scenarios}c);
	\item	STAR – targets are alternating between a fixed point and randomly chosen points inside a square / cube (\figref{fig:scenarios}d);
	\item	EDGE – random distribution over the edge of a square (2D) / surface of a cube (3D), with new targets always being on a different edge (2D) / face (3D) (\figref{fig:scenarios}e);
\end{itemize}

These scenarios can also be treated as test cases of popular applications of UAVs: random and edge scenarios are models of simple open space traffic, line and cross scenarios can be imagined as narrow streets surrounded by urban buildings, while the star scenario is a useful test for drone delivery from a central area to multiple clients at different locations.

\subsection{General Interaction Terms}

Throughout our paper we assume the following realistic conditions~\cite{vasarhelyi2014outdoor}, which highly increase the complexity of the treated problem and makes our work unique even among decentralized approaches:

\begin{itemize}
	\item agents calculate their position information with noise that resembles typical GPS noise;
	\item agents broadcast their position and velocity to other agents at a given framerate $f^{\textrm{s}}$, within a given communication range $r^{\textrm{c}}$, with a certain communication delay $t^{\textrm{del}}$;
	\item agents have maximum velocity $v^{\textrm{max}}$ and inertia, i.e., maximal acceleration $a^{\textrm{max}}$.
\end{itemize}

In our agent-based model we use two general interaction terms: a linear force law for short-range repulsion and a viscous-friction like term for velocity alignment. Repulsion and alignment was first used to establish and maintain correlated collective motion~\cite{reynolds1987flocks},~\cite{vicsek1995novel}, but they have also been used to increase stability of collision-free formation flights and collective target tracking~\cite{vasarhelyi2014outdoor}. In the current model, agents might have individual target points, thus using alignment to increase velocity correlation could seem counter-intuitive. However, besides synchronizing motion, viscous friction also dampens delay-induced oscillations and relaxes high-speed frontal encounters which are necessary features of collision-free self-organization under realistic conditions.
	
Our current repulsion term defining a desired output velocity component at a given time instant is as follows:

\begin{equation}
\tilde{\bm{v}}_i^{\textrm{rep}}=v^{\textrm{rep}}\sum_{j \in J_i}{S(|\bm{x}_j-\bm{x}_i|, r_0, \gamma^{\textrm{rep}})\cdot\frac{\bm{x}_i-\bm{x}_j}{|\bm{x}_i-\bm{x}_j|}}, 
\label{eq:repulsion}
\end{equation}

where $x_i$ denotes the position vector of the $i$th agent, $J_i$ represents the set of indices where $|\bm{x}_i - \bm{x}_j| < r^c$, $v^\textrm{rep}$ is the strength of the interaction, $r_0$ and $\gamma^\textrm{rep}$ are the range and decay length of the interaction. The spatial part of the interaction $S(\Delta x, r_0, \gamma)$ is based on a sigmoid function with a smooth sinusoidal decay from $r_0-\gamma$ to $r_0$:

\begin{multline}
S(\Delta x, r, \gamma)=\\
\begin{cases}
  1 & \text{if ${\Delta x}$ < ${r - \gamma}$}, \\
  {\frac{1}{2}} \left(1 -cos \left(\frac{\pi}{\gamma} (\Delta x - r) \right) \right) & \text{if ${r - \gamma}$ < ${\Delta x}$ < r}, \\
  0 & \text{otherwise.} 
\end{cases}
\label{eq:sigmoid}
\end{multline}

The alignment term is as follows:

\begin{multline}
 \tilde{\bm{v}}_i^{\textrm{frict}}=C^{\textrm{frict}} \sum_{j \in J_i}{ \Big (S(|\bm{x}_j-\bm{x}_i|, r_0+d, \gamma^{\textrm{frict}})} \cdot \\
 \cdot\left(\frac{|\bm{v}_j-\bm{v}_i|}{v^\textrm{ref}}\right)^{\alpha^\textrm{frict}}\cdot\left(\bm{v}_j-\bm{v}_i\right) \Big), 
\label{eq:friction}
\end{multline}

where $v_i$ denotes the velocity vector of the $i$th agent, the interaction is parametrized with $\gamma^{\textrm{frict}}$ decay length and $d$ relative interaction range. The overall strength of the term is defined by $C^\textrm{frict}$, $\alpha^\textrm{frict}$ sets the dependence on the velocity difference between agents $i$ and $j$, and ${v^\textrm{ref}}$ is a velocity scaling parameter which guarantees that the quantity inside the power function is dimensionless. Note that the term kind of resembles viscous friction if $\alpha=0$, however, here we allow for a more general dependence on the velocity difference.

Both interaction terms have a maximum threshold velocity magnitude ($v_{\textrm{max}}^{\textrm{rep}}$ and $v_{\textrm{max}}^{\textrm{frict}}$):

\begin{equation}
\bm{v}_i^{\textrm{rep/frict}}=\\
\begin{cases}
 \tilde{\bm{v}}_i^{\textrm{rep/frict}} & \text{if $|{\tilde{\bm{v}}_i^{\textrm{rep/frict}}}|$ < $v_{\textrm{max}}^{\textrm{rep/frict}}$}, \\
 v_{\textrm{max}}^{\textrm{rep/frict}} \frac{\tilde{\bm{v}}_i^{\textrm{rep/frict}}}{{|\tilde{\bm{v}}_i^{\textrm{rep/frict}}|}} & \text{otherwise.}
\end{cases}
\end{equation}

\begin{figure}[thpb]
	\centering
  \includegraphics[scale=0.7]{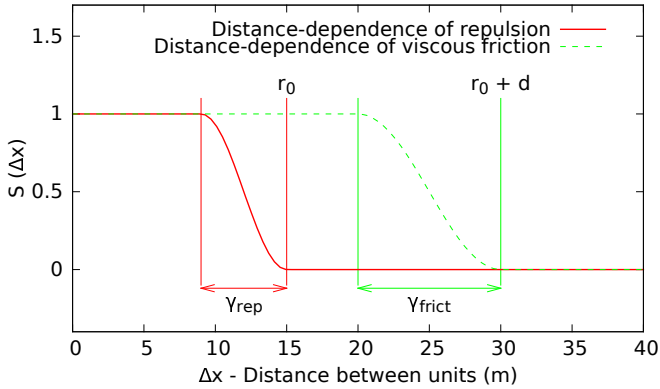}
	\caption{Distance-dependence of the general pairwise interaction terms with specific interaction ranges and decay shapes characterized by $\gamma^{\textrm{rep}}$, $r_0$, $\gamma^{\textrm{frict}}$ and $d$.}
	\label{fig:interactions}
\end{figure}

Since our goal is to reach a target point, we need a driving velocity term always pointing towards our current target point:

\begin{equation}
\tilde{\bm{v}}_i^\textrm{target}=v_0\frac{\bm{x}_i^\textrm{target} - \bm{x}_i}{|\bm{x}_i^\textrm{target} - \bm{x}_i|},
\end{equation}

where $v_0$ is the preferred common travelling speed of all agents and $\bm{x}_i^\textrm{target}$ is the actual target point of agent $i$. For multirotor-type aircraft we implement a velocity decay in the vicinity of the target:

\begin{equation}
\bm{v}_i^\textrm{target} = \tilde{\bm{v}}_i^\textrm{target}D\left(|\bm{x}_i^\textrm{target} - \bm{x}_i|, a, p, v_0\right), 
\end{equation}

where the $p$ gain determines the crossover point between the two phases of deceleration, $a$ is the preferred acceleration and $D(.)$ is a smooth velocity decay function in space, with constant acceleration at high speeds and exponential approach in space at low speeds~\cite{meier2011pixhawk}:

\begin{multline}
D(x, a, p, v_0)=\\
\begin{cases}
  \min (xp,v_0) & \text{if $xp<a/p$}, \\
  \min (\sqrt{2ax - a^2/p^2},v_0) & \text{otherwise.} 
\end{cases}
\end{multline}

In our constant velocity model we cannot slow agents down near a target point thus we simply set $\bm{v}_i^\textrm{target} = \tilde{\bm{v}}_i^\textrm{target}$.

\subsection{Constant Velocity Algorithm}

The Constant Velocity (CV) Algorithm keeps the desired \textit{speed} of agents at all times and prevents collisions by using the repulsion and alignment terms only:

\begin{equation}
\bm{v}_i^\textrm{desired} = v_0\frac{\bm{v}_i^\textrm{rep} + \bm{v}_i^\textrm{frict} + \bm{v}_i^\textrm{target}}
{|\bm{v}_i^\textrm{rep} + \bm{v}_i^\textrm{frict} + \bm{v}_i^\textrm{target}|}
\end{equation}

\subsection{First-in First-out Algorithm}

\begin{figure}[thpb]
	\centering
  \includegraphics[scale=0.5]{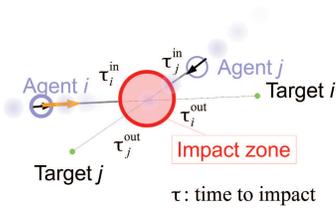}
	\caption{Illustration of the FIFO algorithm. We give yield, i.e. reduce our speed if we assume to arrive to a trajectory cross section (impact zone) after the other agent enters, but before it leaves the impact zone.}
	\label{fig:slowdown}
\end{figure}

The First-in First-out (FIFO) algorithm tends to keep the \textit{direction} of agents more and prevents collisions by slowing down agents that seem to arrive to a future impact zone later than others (\figref{fig:slowdown}). The algorithm consists of the following steps for a given agent $i$:

\begin{enumerate}
\item For every agent $j$ we calculate the angle $\sin{\alpha_{ij}} = \frac{|\bm{v}_i\times\bm{v}_j|}{|\bm{v}_i||\bm{v}_j|}$, the closest points and the distance $d_{ij}$ between the linearly extrapolated trajectories of $i$ and $j$, and check if we reach these closest points within a trajectory extrapolation time $\tau^\textrm{fifo}$. In 2D, the closest point is the intersection point and $d_{ij}=0$, in 3D $d_{ij}$ can be positive and thus there are separate closest points on the two trajectories. We let the general interaction terms handle parallel trajectory cases ($\sin{\alpha_{ij}}=0$) and exclude them here.

\item We define the impact zone radius $r^\textrm{zone}_{ij}$ around the closest points as the largest distance inside which agents $i$ and $j$ could be closer to each other than a predefined distance $r^{\textrm{fifo}}$:
	
\begin{equation}
	r^\textrm{zone}_{ij} = \frac{\sqrt{{r^{\textrm{fifo}}}^2 - d_{ij}^2}}{\sin{\alpha_{ij}}},
\end{equation}
	
assuming that $d_{ij} \le r^{\textrm{fifo}}$ (otherwise we do not use FIFO interaction).

\item We calculate the time needed to arrive to ($\tau^{in}_i$) and leave ($\tau^{out}_i$) the impact zone. If there is any agent $j$ for which $\tau^{in}_i \in [\tau^{in}_j, \tau^{out}_j]$, we reduce our actual target velocity to $\hat{\bm{v}}^\textrm{target}_i = \alpha^\textrm{fifo}\bm{v}^{\textrm{target}}_i$, where $\alpha^\textrm{fifo} \in [0,1]$, otherwise $\hat{\bm{v}}^\textrm{target}_i = \bm{v}^\textrm{target}_i$. 

\end{enumerate}

The final desired velocity of the FIFO algorithm is as follows:

\begin{equation}
\bm{v}_i^\textrm{desired} = \bm{v}_i^\textrm{rep} + \bm{v}_i^\textrm{frict} + \hat{\bm{v}}_i^{\textrm{target}}
\end{equation}

Note that the FIFO algorithm can possibly slow down agents way before they reach a repulsive zone thus it can keep original trajectories unchanged. On the other hand, this model also allows for the case when $\bm{v}_i^\textrm{desired} > \bm{v}_i^\textrm{target}$, when repulsive forces are present.
 
\subsection{Realistic conditions}

Both algorithms were tested in our simulation framework under realistic conditions, with fixed environmental parameters. Noise and delay were artificially added to the position and velocity of other agents an agent obtained at a given time instant. Interactions were spatially localized with limited communication range (agents outside ones communication range were excluded from any local calculations at every moment). Finally, finite acceleration and a general exponential PID-model determined the real velocities of agents from the desired output velocity $v^\textrm{desired}$ calculated by the algorithms and fed to the main differential equation of the framework. The environmental parameters are summarized in Table \ref{table:realistic_conditions}. 

\begin{table}[h]
\caption{Parameters of the realistic setup}
\label{table:realistic_conditions}
\begin{center}
\begin{tabular}{|c|c|c|}
~~~ parameter ~~~ & ~~ value ~~ & ~~ unit ~~ \\
\hline
$v_0$ & 4 & m/s \\
$v^{\textrm{max}}$ & 8 & m/s \\
$a^{\textrm{max}}$ & 6 & $\textrm{m/}\textrm{s}^2$ \\
$p$ & 0.5 & $1/\textrm{s}$ \\
$t^{\textrm{del}}$ & 1 & s \\
$r^{\textrm{c}}$ & 80 & m \\
$f^{\textrm{s}}$ & 5 & Hz
\end{tabular}
\end{center}
\end{table}

\section{EVOLUTIONARY OPTIMIZATION}

As we saw in the last section, the necessity of several interaction terms under realistic conditions resulted in a substantial number of parameters which might have to be varied optimally for each specific situation. Our intention was to provide very generic models and then perform heuristic optimization on the models to find stable and optimal solutions. We used the CMA-ES algorithm~\cite{hansen2003reducing} as a state-of-the-art evolution strategy for continuous parameter optimization to fine-tune our models in all scenarios. Note that the evolutionary optimization was performed at a meta level, i.e., the population consisted of simulations with different parameters, not agents themselves. To decide whether a selected parameter setup works well, we defined a single fitness function, consisting of two main parts:

\begin{itemize}
	\item Agents should not collide with each other. Collision occurs if two agents are closer to each other than $r^{\textrm{coll}}$.
	\item Agents should have high effective velocity, i.e., their velocity vector must point towards their target points as much as possible.
\end{itemize}

To take the first part into account, we define the collision risk:

\begin{equation}
\psi^{\textrm{coll}} (t) = \frac{1}{N(N - 1)} \sum_{i=1}^{N} \sum_{j \neq i} \theta(r^{\textrm{coll}}  - |\bm{x}_i (t) - \bm{x}_j (t) |),
\end{equation}

where $\theta(.)$ is the Heaviside step function. The collision risk is non-zero if any collision occurs at time $t$.

To take the second part into account, we define the effective velocity as a signed projection of the velocity vector onto the line which connects the last ($\bm{x}_i^{\textrm{target}, \textrm{last}}$) and next ($\bm{x}_i^{\textrm{target}, \textrm{next}}$) target points:

\begin{multline}
v^{\textrm{eff}} (t) = \bm{v}_i (t) \left( \frac{\bm{x}_i^{\textrm{target}, \textrm{next}} - \bm{x}_i^{\textrm{target}, \textrm{last}}}{|\bm{x}_i^{\textrm{target}, \textrm{next}} - \bm{x}_i^{\textrm{target}, \textrm{last}}|} \right) \cdot \\
\cdot \begin{cases} 
1 & \text{ if $(\bm{x}_i^{\textrm{target},\textrm{next}} - \bm{x}_i)(\bm{x}_i^{\textrm{target}, \textrm{next}} - \bm{x}_i^{\textrm{target}, \textrm{last}}) > 0$,} \\
-1 & \text{ otherwise.}
\end{cases}
\end{multline}

The second part of this product provides an additional negative sign if the projected position of the agent passes the projected position of its target point.

With the time average of these two parameters, we can define a fitness function:

\begin{equation}
F = \frac{A^2}{(<\psi^{\textrm{coll}}>_t + A)^2} \theta \left( \frac{<v^{\textrm{eff}}>_t}{v_0} \right) \frac{<v^{\textrm{eff}}>_t}{v_0},
\end{equation}

where $A$ sets the tolerance level of the collisions. We aim for low collision risk and high effective velocity simultaneously: in that case, the value of $F$ is near $1$. For declaring a very harsh fitness criterion guaranteeing a significant drop of $F$ even if a single collision occurs with $N=100$, we choose now $A=0.000002$. Note that our intention in the first round was not to eliminate collisions with zero-tolerance but to compare the risk of collisions in different scenarios and densities. However, the collision risk tolerance can be further reduced with smaller $A$ or with executing evolution using larger $r^\textrm{coll}$.

The optimization was performed using our realistic simulation framework~\cite{viragh2014flocking} on the Atlasz supercomputer cluster of the Eötvös University, Budapest, Hungary~\cite{atlasz}, with a population size of 100 simulations and maximum 100 generations (which turned out to be sufficient in all cases). Length of simulations was defined as the time needed for an agent to travel 10 times the characteristic size $L$ of the scenario. Simulations were executed with 100 agents, with random initial placement, agents not being closer to each other than 6 m. In the case of cross and line scenarios, initial placement was limited to 20 m distance from the lines defined by the target points to eliminate false transient free motion in open space at the beginning. Statistical outputs of the first 30 s were also neglected in each scenario to exclude initial transients. Execution time of a single simulation was typically 2 times faster than real time.

Since evolutionary optimization takes a long time, we have chosen typical densities subjectively for each scenario where the mean free path of the agents was neither too short nor too long (i.e. there were many trajectory-conflicts but the system did not yet get jammed).
Our definition of the mean free path is defined as the average linear distance between $N$ evenly distributed agents in a $Dim$-dimensional arena with linear size $L$:

\begin{equation}
MFP = L/N^\frac{1}{Dim}
\end{equation}

Note that for the line and cross scenarios agents typically do not spread out in the whole space available, therefore, this mean free path definition results there in a locally denser situation along the lines of motion (that is why we have chosen larger $MFP$ values for the evolution in these scenarios). However, to be able to compare results in different scenarios, we decided to keep this universal density definition at all times.

\begin{figure}[thpb]
	\centering
  \includegraphics[width=0.45\textwidth]{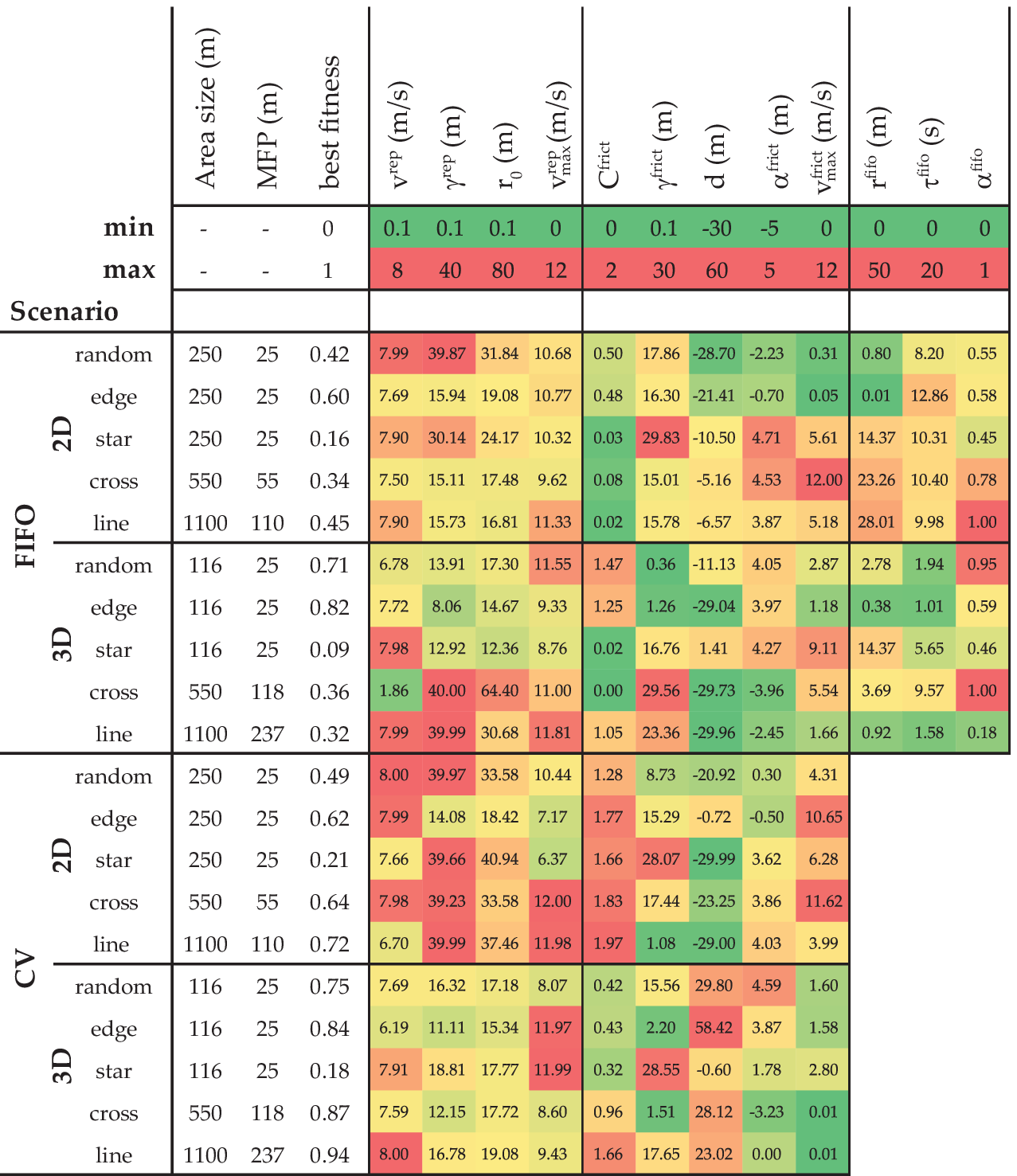}
	\caption{Parameter setup of all models in all scenarios in 2D and 3D as a best-fitness result of the CMA-ES optimization. Color coding from green through yellow to red represents the relative value of a given parameter within its defined maximal range.}
	\label{fig:evol_params}
\end{figure}

The optimized parameter values for each scenario are summarized in \figref{fig:evol_params}.


Note that we performed only one evolution for each scenario with 10k fitness-evaluations each, but running the same evolution many times could lead to different results, i.e., different local optima. However, we can gain significant knowledge even from this single evaluation. 
\begin{itemize}
	\item In general flocking models, the typical range of alignment is usually larger than the typical range of repulsion, i.e., alignment functions at a distance where there are already no hard-core repulsive forces. Contrarily, in these traffic models the same alignment term has an effective range that is always below the range of repulsion. As a consequence, alignment here serves as an additional helper to reduce collisions and, more significantly, delay induced oscillations.
	\item In the FIFO model, the slow-down and alignment terms have overlapping function. In the random and edge scenarios, where each agent has different target points, the alignment and the slow-down terms are basically not existing or very small, collisions are mostly handled by repulsion. In the other scenarios, where jammed states around a common target point occur, which is a typical excitation for oscillations when delays are present, both terms are significant and are definitely needed.
\end{itemize}

\section{COMPARISON OF MODELS}

We measured the effective velocity and the collision risk of each optimized model in each scenario in 2D and 3D environment as a function of agent density. We used the following flux definition to evaluate the overall throughput of the models:

\begin{equation}
\Phi = \frac{v^\textrm{eff}}{MFP}.
\end{equation}

\begin{figure*}[thpb]
	\centering
	\includegraphics[scale=0.8]{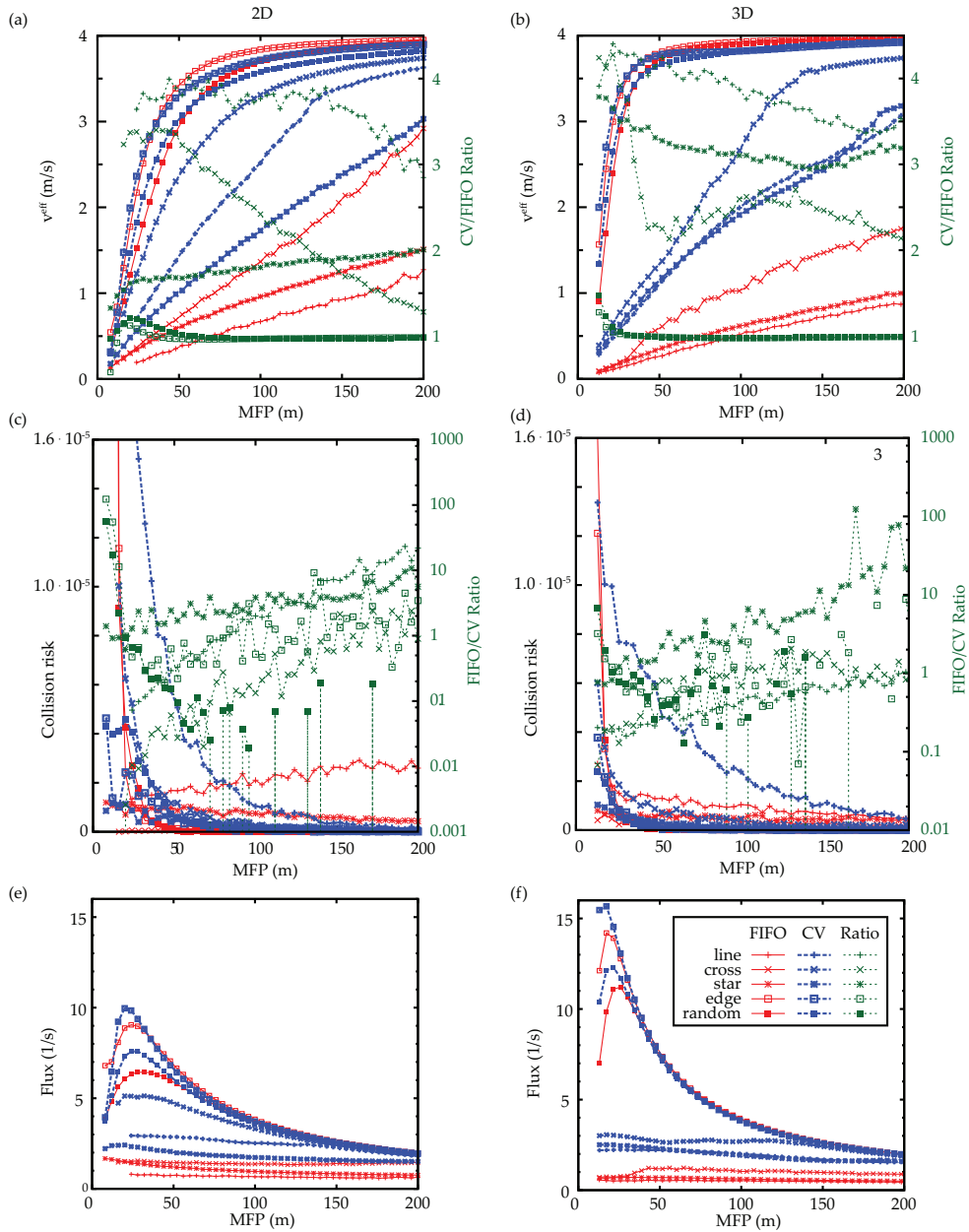}
	\caption{Performance of the FIFO and CV algorithm in different densities in 2D (left column) and 3D (right column) space. a-b) effective velocity, as a function of mean free path. Decreasing the MFP (i.e. increasing density) also decreases $v^\textrm{eff}$. Ratio of the effective velocities of the two models indicates that in general the constant velocity model can yield higher efficiency; however, results are scenario-specific; c-d) collision risk as a function of mean free path. Collision risk nearly always increases with density. Ratio of the collision risk of the two models indicates that best model selection depends on density and scenario; e-f) average flux as a function of mean free path. The random and edge models show a typical flux distribution, where flux is maximal at the critical density point between free motion and the  jammed state.}
	\label{fig:results}
\end{figure*}

All density-scan results are visualized in \figref{fig:results}. It is clear in all cases that increasing the agent density decreases the effective velocity and also increases the probability of collisions. However, the flux also increases with density even up to the stage where the effective velocity is significantly reduced, and then, for some models it eventually drops. The transition from the dynamic to the jammed state as a function of density seems to be a smooth one in all models; however, typical transition diagrams between the two states with a maximal flux around a mid-way critical density can only be observed in the random and edge cases. In the other scenarios, according to the visual observation of the actual simulations, the jammed state around common target points are never really eliminated thus no true transition can occur there.

The fact that the maximal flux increases with reduced effective velocity at higher densities reflects the so called slower-is-faster (SIF) effect~\cite{GershensonHelbing2015SIF}. Even though we do not explicitly reduce the target velocity here to increase the flux, in the FIFO model, we do reduce the average speed implicitly by increasing the number of trajectory interferences and thus the amount of the slowdown behaviour. However, for a clear investigation of the SIF effect more simulations would be needed with changing $v^\textrm{target}$.

In general, each situation has its own challenges, and none of the algorithms was capable of solving all problems with the same efficiency. In fact, there are three main challenges in the used scenarios: 

\begin{itemize}
  \item handle conflicting trajectories in general (random, edge, and all models to some extent)
	\item handle fixed junction points and kind-of-fixed lanes with self-organization and possibly emergent behaviour (cross, line);
	\item handle queuing at a common target point (cross, line, star).
\end{itemize}	
	
In fact, the third problem is not treated explicitly in any of the presented models and thus resolving these situations is the least efficient in all cases. An explicit solution would require additional communication between agents to, e.g., discuss an order and patiently stand in a queue and give yield accordingly.

The FIFO algorithm was designed and is most suitable for random/edge situations where individuals have different targets, i.e., in open space general traffic. However, with the currently evolved parameters it fails with high collision risk at high densities, since oscillations emerge due to the communication delay and the strong repulsion. It also fails at situations with a common target, since agents with reduced speed are not able to decide on an appropriate order and get completely jammed. Note that using, e.g., the star-evolved parameters with large friction and slowdown effect for the random/edge situations gives significantly better results in terms of collision risk but reduce the overall effective velocity. A further investigation would be to evolve the model using all scenarios simultaneously to find a globally optimal parameter selection.

The CV algorithm can handle those situations significantly better (in terms of effective velocity or flux), where multiple agents have the same target point, since the constant velocity constraint works against jamming, however its solution is still in the jammed phase and could be further enhanced with explicit queuing rules. Regarding collision risk there is no clear global best model - performance depends on both density and selected scenario. 

Interestingly the 2D and 3D scenarios give very similar outcome in terms of the effective velocity, flux and number of collisions, which means that the three dimensional models are capable of handling a lot more UAVs in the air above a given area.

\section{DISCUSSION}

One interesting aspect of the used models is that they can produce different emergent phenomena in special situations:

\begin{itemize}
	\item Self-organized lane formation is present in the line and cross scenarios for both models. This structure with optimal parameter space selection for maximizing traffic flux has already been observed in army ants~\cite{couzin2003self} and pedestrians~\cite{helbing2001self};
	\item self-excited oscillations as a result of communication delay, such as in ghost traffic jams on highways, are very common in all cases, especially at high densities and with common targets.
\end{itemize}

In fact, delay turns out to be the most important environmental constraint, with a fundamental effect on the overall flux and more especially on the number of collisions (through exciting oscillations). We performed further simulations to reveal the effects of the communication delay (see Table \ref{table:delay}). Removing the 1s communication delay from the environment resulted in an average 90\% of reduction of collisions and over 20\% increase in the effective velocity based on simulations executed with the parameters used for evolutionary optimization in each situation. This has two different messages; in one hand, decreasing communication delays in real life robotic systems should be a main technological improvement, and, in the other hand, performing simulations without the presence of delays could lead to false and unstable results in real environments.

\begin{table*}[]
\centering
\caption{Effects of the communication delay on effective velocity and collisions}
\label{table:delay}
\begin{tabular}{|l|r|r|r|r|r|r|r|r|}
& \bfseries MFP &\IEEEeqnarraymulticol{2}{t}{\textbf{$v^\textrm{eff}$}} && 
\IEEEeqnarraymulticol{2}{t}{\textbf{collisions}} && \textbf{decrease}\\
& &$t^{\textrm{del}}=0s$&$t^{\textrm{del}}=1s$&$v^\textrm{eff}_{0s}/v^\textrm{eff}_{1s}$&
$t^{\textrm{del}}=0s$&$t^{\textrm{del}}=1s$&$\textrm{coll}_{0s}/\textrm{coll}_{1s}$ & \textbf{of coll.}\\
\textbf{situation} & (m) & (m/s) & (m/s) & ratio & & & ratio & 1s $\to$ 0s\\
\hline
FIFO 2D random  & 25   & 2.04  & 1.66  & 1.23  & 0.65  & 3.15   & 0.21  & 79\%   \\
FIFO 2D star    & 25   & 0.70  & 0.64  & 1.10  & 0.80  & 1.60   & 0.50  & 50\%   \\
FIFO 2D edge    & 25   & 2.68  & 2.33  & 1.15  & 0.05  & 3.40   & 0.02  & 99\%   \\
FIFO 2D cross   & 55   & 1.07  & 0.88  & 1.22  & 0.15  & 1.20   & 0.13  & 88\%   \\
FIFO 2D line    & 110  & 1.37  & 1.08  & 1.27  & 0.15  & 13.70  & 0.01  & 99\%   \\
CV  2D random  & 25   & 2.46  & 1.92  & 1.28  & 0.00  & 4.30   & 0.00  & 100\%  \\
CV  2D star    & 25   & 1.10  & 0.78  & 1.42  & 0.25  & 2.00   & 0.13  & 88\%   \\
CV  2D edge    & 25   & 2.77  & 2.46  & 1.13  & 0.00  & 1.55   & 0.00  & 100\%  \\
CV  2D cross   & 55   & 2.95  & 2.51  & 1.18  & 0.00  & 2.80   & 0.00  & 100\%  \\
CV  2D line    & 110  & 3.49  & 2.74  & 1.27  & 0.05  & 4.80   & 0.01  & 99\%   \\
\hline
\textbf{average} & & & & \textbf{1.22} & & & \textbf{0.10} & \textbf{90\%}
\end{tabular}
\end{table*}

The finite communication range as an environmental limitation does not seem to be a severe bottleneck at all, since all interactions are local and decay quickly with distance anyway. Using unlimited communication did not result in better solutions in general, but of course, communication range has to be scaled with speed, for example. The fact that general 3D traffic can be based on pure local interactions is highly advantageous when we need to scale up system size for global air-traffic control without increasing complexity or costs of infrastructure.

In summary, we have shown two distinct models which are tunable to handle realistic dense UAV traffic scenarios in two and three dimensions. The FIFO model is mostly successful for open space traffic, while the CV model is more efficient to handle situations with a common target point. The models were tested in realistic situations, where communication delay turned out to be the most important factor regarding the number of induced collisions and thus destabilization of motion. Further investigations would be needed to generalize the models to higher and heterogeneous preferred velocity ranges and to provide a global optimization that includes selective behaviour for more than one scenarios simultaneously that could possibly occur naturally on the way in a self-organized common airspace.

The scenarios studied were simplistic. Still, they show the potential of developing self-organizing air traffic management systems, where only local, real-time information is required to solve potential navigational conflicts. Certainly, a structured environment can contribute to coordinate air traffic, e.g., with dedicated lanes. However, we envision efficient air traffic scenarios without a central repository of UAV data.

\bibliography{IEEEabrv,uav_traffic}

\addtolength{\textheight}{-12cm}   


\section*{ACKNOWLEDGMENT}

Thanks for Tamás Vicsek for his coordination and useful hints. Also thanks for Gusz Eiben for his keen enthusiasm and help regarding evolutionary optimization.

\section*{DISCLAIMER}

Disclaimer: This work has been accepted for publication atthe IEEE IROS 2016 Conference. Copyright with IEEE. Personal use of this material is permitted. However, permission to reprint/republish this material for advertising or promotional purposes or for creating new collective works for resale or redistribution to servers or lists, or to reuse any copyrighted component of this work in other works must be obtained from the IEEE. This material is presented to ensure timely dissemination of scholarly and technical work. Copyright and all rights therein are retained by authors or by other copyright holders. All persons copying this information are expected to adhere to the terms and constraints invoked by each author’s copyright. In most cases, these works may not be reposted without the explicit permission of the copyright holder. For more details, see the IEEE Copyright Policy.

\end{document}